\documentclass{PoS}

\title{NLO QCD corrections to processes with multiple electroweak bosons}

\ShortTitle{VBFNLO}

\author{\speaker{Dieter Zeppenfeld}%
         \thanks{On behalf of the VBFNLO Collaboration, with contributions 
        from  K.~Arnold, M.~B\"ahr, T.~Figy, N.~Greiner, C.~Hackstein, 
        G.~Kl\"amke, M.~Kubocz, S.~Palmer, S.~Prestel, M.~Rauch, and H.~Rzehak. 
        Support by BMBF under Grant No.~05H09VKG  (``Verbundprojekt HEP-Theorie'') 
        and DFG within SFB/TR9 is gratefully acknowledged. }\\
        Karlsruhe Institute of Technology\\
        E-mail: \email{dieter.zeppenfeld@kit.edu}}

\author{G.~Bozzi\\
Dipartimento di Fisica, Universit\`a di Milano and INFN,
Sezione di Milano, Via Celoria 16, 20133 Milano, Italy
}

\author{ F.~Campanario, C.~Englert,  V.~Hankele, S.~Pl\"atzer\\
         Karlsruhe Institute of Technology\\
         E-mail: \email{vbfnlo@particle.uni-karlsruhe.de}}

\author{B.~J\"ager\\
Institut f\"ur Theoretische Physik und Astrophysik, 
Universit\"at W\"urzburg, 97074 W\"urzburg, Germany	}

\author{C.~Oleari \\
         Universit\`a di Milano-Bicocca and INFN, Sezione di Milano-Bicocca, 
         Piazza della Scienza 3, 20126 Milan, Italy}

\author{M.~Spannowsky\\
Institute for Theoretical Science, 5203 University of Oregon,
Eugene, OR 97403, USA
}

\author{M.~Worek\\
Fachbereich C Physik, Bergische Universit\"at Wuppertal, 
42097 Wuppertal, Germany 
}

\abstract{The VBFNLO program package is a collection of Monte Carlo 
programs for the calculation of NLO QCD corrections to vector boson 
fusion cross sections, double and triple vector boson production, or 
the production of two electroweak bosons in association with an 
additional jet. An overview is given of the processes and features 
implemented in VBFNLO. $WW\gamma$ and $W\gamma j$ production are 
discussed as examples.}

\FullConference{RADCOR 2009 - 9th International Symposium on Radiative 
Corrections (Applications of Quantum Field Theory to Phenomenology) \\
                 October 25-30 2009\\
                 Ascona, Switzerland}

\def\sla#1{\ifmmode%
\setbox0=\hbox{$#1$}%
\setbox1=\hbox to\wd0{\hss$/$\hss}\else%
\setbox0=\hbox{#1}%
\setbox1=\hbox to\wd0{\hss/\hss}\fi%
#1\hskip-\wd0\box1 } 

\begin{document}

\section{Introduction}
\label{sec:intro}

The highest energy particle interactions which can be studied in the 
laboratory are being provided 
by $p\bar p$ collisions at the Fermilab Tevatron and $pp$ collisions at the 
CERN Large Handron Collider (LHC). In order to compare hadron collider 
data with theoretical predictions at a sufficiently precise level, 
NLO QCD corrections are needed for many signal and background processes. 
A sizable fraction of these proceedings is devoted to this topic. 

The purpose of the present contribution is to provide an overview of NLO QCD 
calculations which have been implemented in the publicly available 
VBFNLO~\footnote{The source code for the VBFNLO programs is available at 
http://www-itp.particle.uni-karlsruhe.de/vbfnlo/}  program 
package~\cite{Arnold:2008rz} or which have recently been performed in the 
VBFNLO framework. In addition, we discuss  NLO QCD 
corrections to $WW\gamma$ production~\cite{Bozzi:2009ig} 
in Section~\ref{sec:VVgamma} and to 
$W\gamma j$ production~\cite{Campanario:2009um} in Section~\ref{sec:WgammaJet}.

\section{Overview of VBFNLO Processes}
\label{sec:processes}

The VBFNLO package provides parton level Monte Carlos for the calculation
of hadron collider cross sections and distributions with NLO QCD accuracy.
The cancellation of collinear and soft divergences between virtual 
contributions and real emission corrections is achieved with the 
dipole subtraction method of Catani and Seymour~\cite{Catani:1996vz}. 
The calculation of matrix elements is based on the helicity amplitude 
formalism of Ref.~\cite{Hagiwara:1985yu}, while dedicated routines 
using Passarino-Veltman and/or Denner-Dittmaier 
recursion~\cite{Passarino:1978jh} are providing numerical
tensor reduction and the calculation of the finite parts of virtual amplitudes.

VBFNLO originates from vector boson fusion (VBF) processes which can be 
pictured as quark-(anti)quark scattering via $t$-channel electroweak 
boson exchange, with the emission of additional weak bosons or a Higgs 
boson off the quark- or electroweak boson lines. All processes, which are
discussed below,  
neglect identical fermion effects. For VBF processes such as 
$q\bar Q \to q\bar Q H$ this implies that $s$-channel diagrams 
such as $u\bar u\to ZH\to u\bar u H$ are considered to be separate processes
which are not provided by VBFNLO. Indeed, in the phase space regions where VBF
processes can be isolated at the LHC, these $s$-channel contributions are 
usually small and their interference with the VBF diagrams is truly 
negligible (see e.g. Ref.~\cite{Bredenstein:2008tm}). 
The following VBF processes at NLO QCD are implemented in the 2008 
release of VBFNLO~\cite{Arnold:2008rz}:
\begin{itemize}
\item $\bf Hjj$ production with and without decay of the Higgs 
boson~\cite{Figy:2003nv}. Options for Higgs decay modes include 
$H\to \tau^+\tau^-$, $H\to W^+W^-\to l^+\nu l^-\bar\nu$, 
$H\to ZZ\to 4$~leptons, $H\to\gamma\gamma$, or 
$H\to b\bar b $. Also, anomalous $HVV$ couplings are 
supported~\cite{Hankele:2006ma}.

\item  $\bf Hjjj$ production with and without decay of the Higgs 
boson~\cite{Figy:2007kv}. Options for Higgs decay modes are the same as 
for $Hjj$ production. The $H+3$-jet calculation does not contain
the full NLO QCD corrections but neglects certain $t$-channel color exchange
contributions which are subleading in $1/N$ and which have been shown to 
be very small for VBF kinematics.

\item $\bf W^\pm jj$ production with subsequent leptonic $W$ 
decay~\cite{Oleari:2003tc}.

\item $\bf Zjj$ production with subsequent leptonic $Z$ 
decay~\cite{Oleari:2003tc}.

\item $\bf W^+W^- jj$ production with subsequent leptonic $W$ decay
and full off-shell effects~\cite{Jager:2006zc}. Anomalous trilinear and 
quartic couplings between Higgs boson and weak bosons are implemented.

\item $\bf ZZ jj$ production with subsequent leptonic $Z$ decay
and full off-shell effects~\cite{Jager:2006cp}. 

\item $\bf W^\pm Z jj$ production with subsequent leptonic weak boson decay
and full off-shell effects~\cite{Bozzi:2007ur}. 

\end{itemize}
For the last three processes ($VVjj$ production in VBF) extra vector 
resonances in $VV$ scattering are implemented within
the context of warped Higgsless models~\cite{Englert:2008wp}.
Beyond the VBF processes listed above, the calculation of 
\begin{itemize}
\item  $\bf W^+W^+jj$ and $\bf W^-W^-jj$ 
production in VBF with leptonic decay of the $W$s 
and full off-shell effects~\cite{Jager:2009xx}
\end{itemize}
has recently been completed and will be made publicly available in a 
future VBFNLO release.

The next large class of processes concerns double and triple electroweak 
boson production with NLO QCD accuracy. At tree level, the underlying reactions
are of the type $q\bar q \to VV$ or $q\bar q \to VVV$. In all cases, 
leptonic decays of the electroweak bosons to lepton pairs and off-shell 
effects are included in 
the calculations. The processes implemented in the 2008 release are 
\begin{itemize}
\item $\bf W^+W^-$ production. This process has been verified against 
MCFM~\cite{Campbell:1999ah}.
\item $\bf W^+W^-Z$ production, including the $H\to WW$ 
resonance~\cite{Hankele:2007sb}.
\item $\bf W^\pm ZZ$ production, including the $H\to ZZ$ 
resonance~\cite{Campanario:2008yg}.
\item $\bf W^\pm W^\mp W^\pm$ production, including the $H\to WW$ 
resonances~\cite{Campanario:2008yg}.
\end{itemize}
Total NLO cross sections for the last three processes ($VVV$ production) have 
been successfully compared against the results of Ref.~\cite{Binoth:2008kt}
which, however, are only available without leptonic decay of the weak 
bosons and which do not include Higgs resonance contributions.

Two extensions of the triple weak boson production processes have recently 
been completed within the VBFNLO framework. These are
\begin{itemize}
\item $\bf W^+W^-\gamma$ and $\bf ZZ\gamma$ production with subsequent 
$W$ and $Z$ leptonic decay~\cite{Bozzi:2009ig}     and
\item $\bf W^\pm\gamma j$ production with $W\to l\nu$ 
decay~\cite{Campanario:2009um}.
\end{itemize}
They will be discussed in more detail below.

\section{$WW\gamma$ Production}
\label{sec:VVgamma}

\begin{figure}[th]
\label{fig:wwafeyn}
\includegraphics[scale = 0.9]{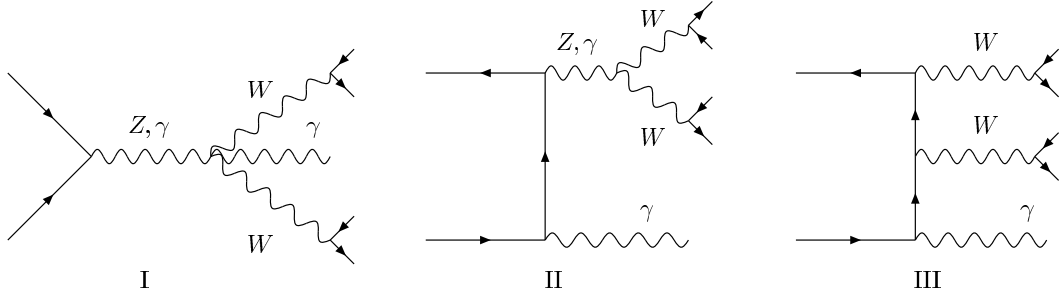}
\caption{Examples of the three topologies of Feynman diagrams
  contributing to $pp\to W^+W^-\gamma+X$.}
\end{figure}

The three classes of Feynman graphs contributing to $WW\gamma$ production 
at tree level are shown in Fig.~\ref{fig:wwafeyn}. As in all VBFNLO 
processes, the decay of the $W$s to charged leptons is included, i.e. the 
QCD corrections are calculated for the full process 
$pp,\; p\bar p\to\nu_1 l_1^+\bar\nu_2 l_2^-\gamma~+X$, including 
final state photon radiation off the charged leptons or other off-shell
contributions. 

Virtual QCD corrections can be considered separately for each of the 
tree level graphs. Corrections to the vertex topology (I) factorize in 
terms of the corresponding Born amplitude, and this includes the soft and 
collinear divergences which appear as $1/\epsilon^2$ and $1/\epsilon$ poles 
in dimensional regularization. The cancellation of these poles against the 
real emission cross sections, where the divergent phase space integral 
is proportional to the full Born amplitude squared, implies that the 
infrared divergent parts of the QCD loop corrections to topologies (II) 
and (III) also factorize in terms of their respective Born amplitude. 
The full virtual contributions are thus given by
\begin{equation}
M_V = \widetilde{M}_V + \ \frac{\alpha_S}{4 \pi} \ C_F \ \left( \frac{4
      \pi \mu^2}{Q^2} \right)^\epsilon  \ \Gamma{(1 + \epsilon)} \ \left[
    -\frac{2}{\epsilon^2} - \frac{3}{\epsilon} - 8 + \frac{4 \ \pi^2}{3}
  \right] \ M_B, \label{eq:MV}
\end{equation}
where $M_B$ is the Born amplitude and $Q$ is the partonic center-of-mass 
energy, i.e. the invariant mass of the final state $WW\gamma$ system,
$m_{WW\gamma}$. The term $\widetilde{M}_V$ consists of the finite parts of 
the virtual 
corrections to 2 and 3 weak boson amplitudes and can be calculated 
numerically in $d=4$ dimensions.

\begin{figure}[b]
\includegraphics[scale = 0.95]{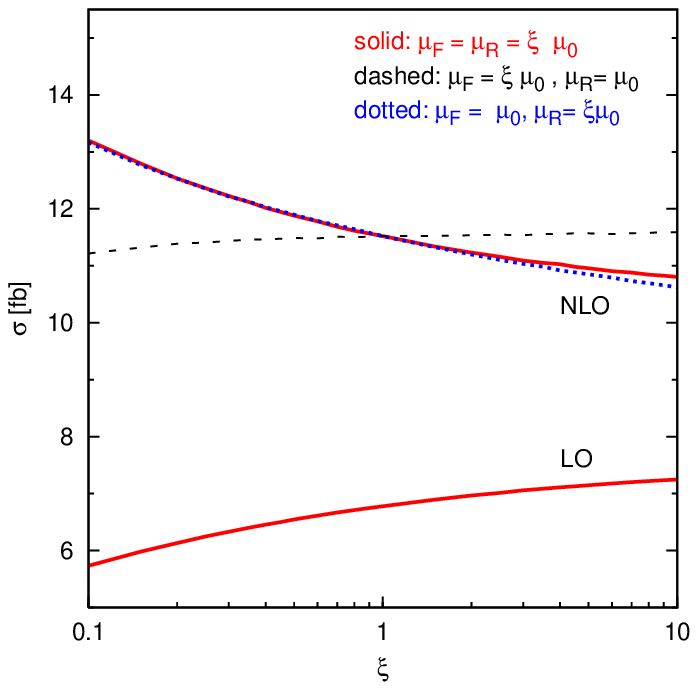}
\hspace{1cm}
\includegraphics[scale = 0.95]{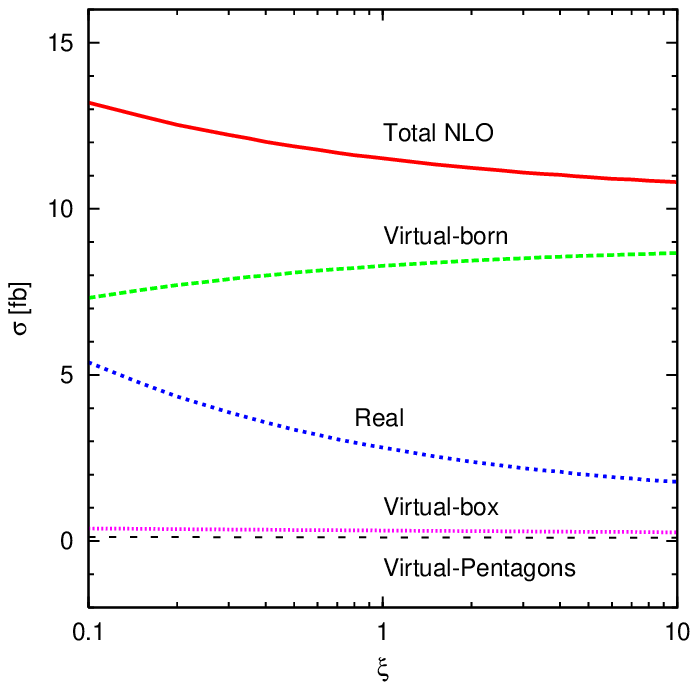}
\caption[]{\label{fig:scale_wwa}
{\it Left:} {Scale dependence of the total LHC cross section for 
  $p p \to W^+W^-\gamma+X \to \ell^+ \ell^- \gamma +\sla{p}_T+X$ at LO and
  NLO within the cuts of Eqs.~(\ref{eq:cuts}) and (\ref{eq:isol}).
  The factorization and renormalization scales are together or
  independently varied in the range from $0.1\mu_0$ to $10\mu_0$, 
  with $\mu_0=m_{WW\gamma}$.} 
{\it Right:} { Same as in the left panel but for the different NLO
  contributions at $\mu_F=\mu_R=\xi\mu_0$.}}
\end{figure}
For our numerical results we  impose a set of minimal cuts on leptons, photon
and jets, namely
\begin{equation}
p_{T_{\gamma(\ell)}} > 20 \ \mathrm{GeV} \qquad
|y_{\gamma(\ell)}| < 2.5 \qquad
R_{\ell\gamma} > 0.4  \qquad
R_{j \ell} > 0.4  \qquad
R_{ j \gamma} > 0.7 
\label{eq:cuts}
\end{equation}
where, in our simulations, a jet is defined as a parton of 
transverse momentum $p_{Tj}>20$~GeV.
For photon isolation, we implement the procedure 
defined in \cite{Frixione:1998jh}: if $i$ is a parton with transverse energy
$E_{T_i}$ and a separation $R_{i\gamma}$ with a photon of transverse momentum
$p_{T_\gamma}$, then for 
\begin{equation}
\Sigma_i \, E_{T_i} \, \theta (\delta - R_{i\gamma}) \, \leq \, p_{T_\gamma} \,
\frac{1-\cos\delta}{1-\cos\delta_0} \,\,\,\,\,\,\,\,\,\,
(\mathrm{for\,\, all} \,\,\,\,\,\delta\leq\delta_0) \label{eq:isol}
\end{equation}
the event is accepted. Here $\delta_0$ is a fixed separation that we 
set equal to 0.7. 

The integrated cross section within these cuts and its scale 
variation around $\mu_0=m_{WW\gamma}$ is depicted in 
Fig.~\ref{fig:scale_wwa}. The scale variation is modest both
at LO and at NLO. The LO factorization scale variation seriously 
underestimates the NLO corrections, which amount to a $K$-factor of 
about 1.7 at the LHC. At NLO, the factorization scale dependence
is substantially reduced when compared to LO. The modest renormalization
scale dependence can be associated to mainly the real emission 
contributions.

\section{$W\gamma j$ Production}
\label{sec:WgammaJet}

\begin{figure}[b]
  \includegraphics[scale = 0.85]{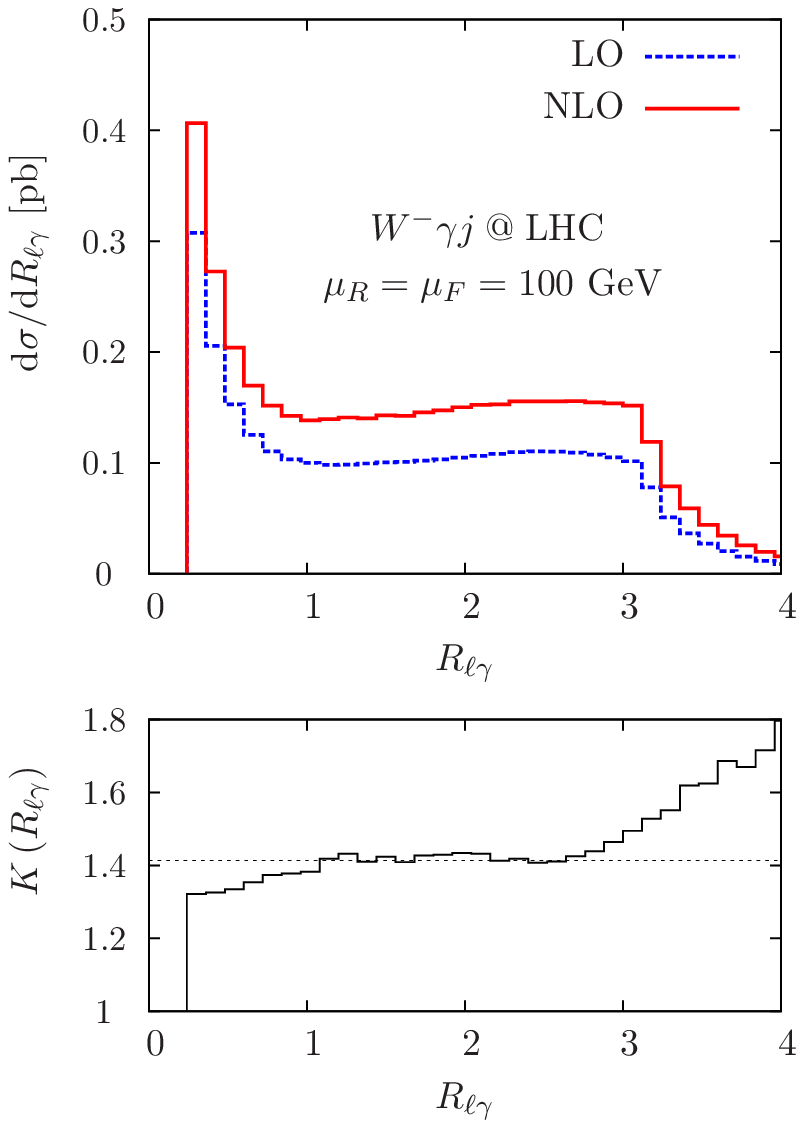}
\hspace{1cm}
  \includegraphics[scale = 0.85]{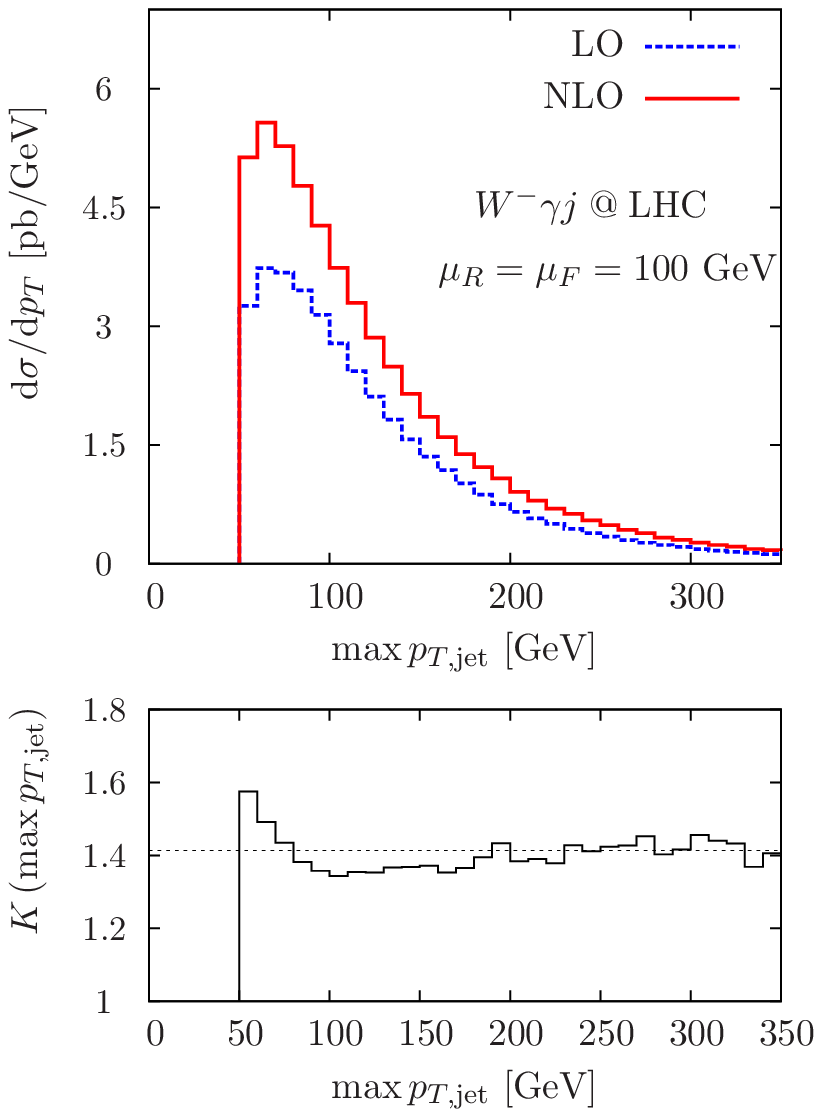}
\vspace{-0.2cm}
\caption{Differential distribution of the photon-lepton separation 
$R_{\ell \gamma}$ and the maximum jet transverse momentum at LO 
(dashed) and at NLO (solid). The lower panels show the differential 
$K$-factor. The dotted lines correspond 
to the $K$-factor of the integrated cross section, $K=1.41$.
\label{fig:waj}
}
\end{figure}

In Ref.~\cite{Campanario:2009um} the NLO QCD corrections to 
$pp,\, p\bar p \to W^\pm\gamma j+X$ cross sections 
have been calculated, again including 
final state photon radiation off the $W$ decay products and finite width 
effects. When varying the factorization and renormalization scales by a 
factor of 2 around fixed values of $\mu_0=100$~GeV one finds modest
scale variations which decrease from about 11\% at LO to 7\% at NLO. 
$K$-factors are around 1.4 at the LHC, but they do vary over phase
space. Two examples for this variation are shown in Fig.~\ref{fig:waj}.

\section{Conclusions}
\label{sec:conclusions}

For a variety of production processes of electroweak bosons, NLO QCD 
corrections have been calculated and implemented in the VBFNLO program
package. QCD corrections are found to be fairly small for VBF cross 
sections.
For triple electroweak boson production, however, QCD corrections increase
LO estimates substantially, by $K$-factors up to 1.8 for integrated cross
sections, and even larger values are observed in certain distributions.
The size of QCD corrections for $W\gamma j$ production falls between
these two extremes. It is clear, however, that NLO QCD corrections 
should be considered for a precise comparison of data to SM predictions
for all these processes.

\end{document}